\documentclass[twocolumn, superscriptaddress, preprintnumbers, titlepage,
letterpaper, amssymb, aps, amsmath]{revtex4} 

\usepackage{epsfig}

\begin{document}

\title{Coarse Grained Microscopic Model of Glass Formers}

\author{Juan P. Garrahan}

\affiliation{Theoretical Physics, University of Oxford, 1 Keble Road,
Oxford, OX1 3NP, U.K.}

\author{David Chandler}

\affiliation{Department of Chemistry, University of California,
Berkeley, California 94720}

\begin{abstract}
We introduce a coarse grained model for atomic glass formers.  Its
elements are physically motivated local microscopic dynamical rules
parameterized by observables.  Results of the model are established
and used to interpret the measured behaviors of supercooled fluids
approaching glass transitions.  The model predicts the presence of a
crossover from hierarchical super-Arrhenius dynamics at short length
scales to diffusive Arrhenius dynamics at large length scales.  This
prediction distinguishes our model from other theories of glass
formers, and can to be tested by experiment.
\end{abstract}

\maketitle

\noindent 
{\bf Introduction.} This article introduces a microscopic model of
glass forming supercooled liquids. It is a coarse grained model with
local dynamical rules, the elements of which can be measured. The
rules contain only the physical features of detailed balance, dynamic
facilitation, and persistence of particle flow direction.  Our
analysis of the model indicates that these features are all that is
required to explain the time dependent and thermal behaviors of
supercooled glass forming liquids. For such systems, relaxation times
grow rapidly with decreasing temperature $T$, usually more rapidly
than exponentially in $T^{-1}$. This non-Arrhenius (or
super-Arrhenius) temperature dependence often coincides with a
relatively large change in heat capacity at the glass transition. On
the other hand, liquids whose relaxation times grow in an Arrhenius
fashion with decreasing temperature often display small changes in
heat capacity at the glass transition.  Apparent correlations like
these between thermodynamic properties and relaxation times have led
to a commonly held opinion that a finite temperature thermodynamic
transition somehow underlies a glass transition (for a review of this
and other views, see \cite{Review}).  In contrast, our model explains
the empirically observed correlations of relaxation and calorimetric
properties at the glass transition without recourse to unusual or
precipitous thermodynamics.

Our model is described in terms of a grid or lattice in space-time.
Underlying this discrete picture are certain ideas about continuous
atomic motions in a supercooled liquid.  The first idea is that
particle mobility is sparse.  Most atomic motions are small amplitude
vibrations and not diffusion steps.  Most molecules are effectively
jammed.  The second idea is that particle mobility in a glass former
is the result of facilitation \cite{Glarum,Fredrickson-Andersen}.
Specifically, a region of jammed atoms can become unjammed and thus
exhibit mobility only when it is adjacent to a region that is already
unjammed.  This idea implies that mobile or unjammed particles form
chains in space-time. Chains of mobile particles have been revealed by
atomistic computer simulations of supercooled liquids
\cite{MDdirectionality}.  The third idea is that mobility carries a
direction, and it is possible for this direction to persist in a glass
former for significant periods of time.  This convective-like quality
of particle motion is also revealed by the atomistic computer
simulations of supercooled liquids \cite{MDdirectionality}.

To help illustrate how these ideas motivate a lattice model, Fig.\
1(a) depicts a mostly jammed fluid of spheres (drawn in two dimensions
for artistic convenience).  It shows, schematically, how mobility is
found in unjammed regions, and that directed motion in one of those
regions might make similarly directed motion possible in adjacent
regions.  The arrows are drawn pointing parallel to the mean direction
of facilitation, and anti-parallel to the mean direction of particle
motion.  Given that particles singly occupy finite volumes, the
possible influence of facilitation falls in a cone, as indicated by
the dashed lines in Fig.\ 1(a).

The point of each cone in Fig.\ 1(a) depicts a position that is made
accessible to an atom by prior motion in the vicinity of that point.
As large voids dissipate quickly in a dense material, the most likely
motions are collective and complicated shuffling of several particles.
This shuffling could have a number of causes.  One possibility is the
presence of local density that is slightly less than a constraining
value.  Whatever its precise nature, the collective motion will have a
definite direction for a period of time.  In highly networked systems,
directional persistence will quickly dissipate, decreasing its effect
on subsequent motion.

\begin{figure}[ht]
\begin{center}
\epsfig{file=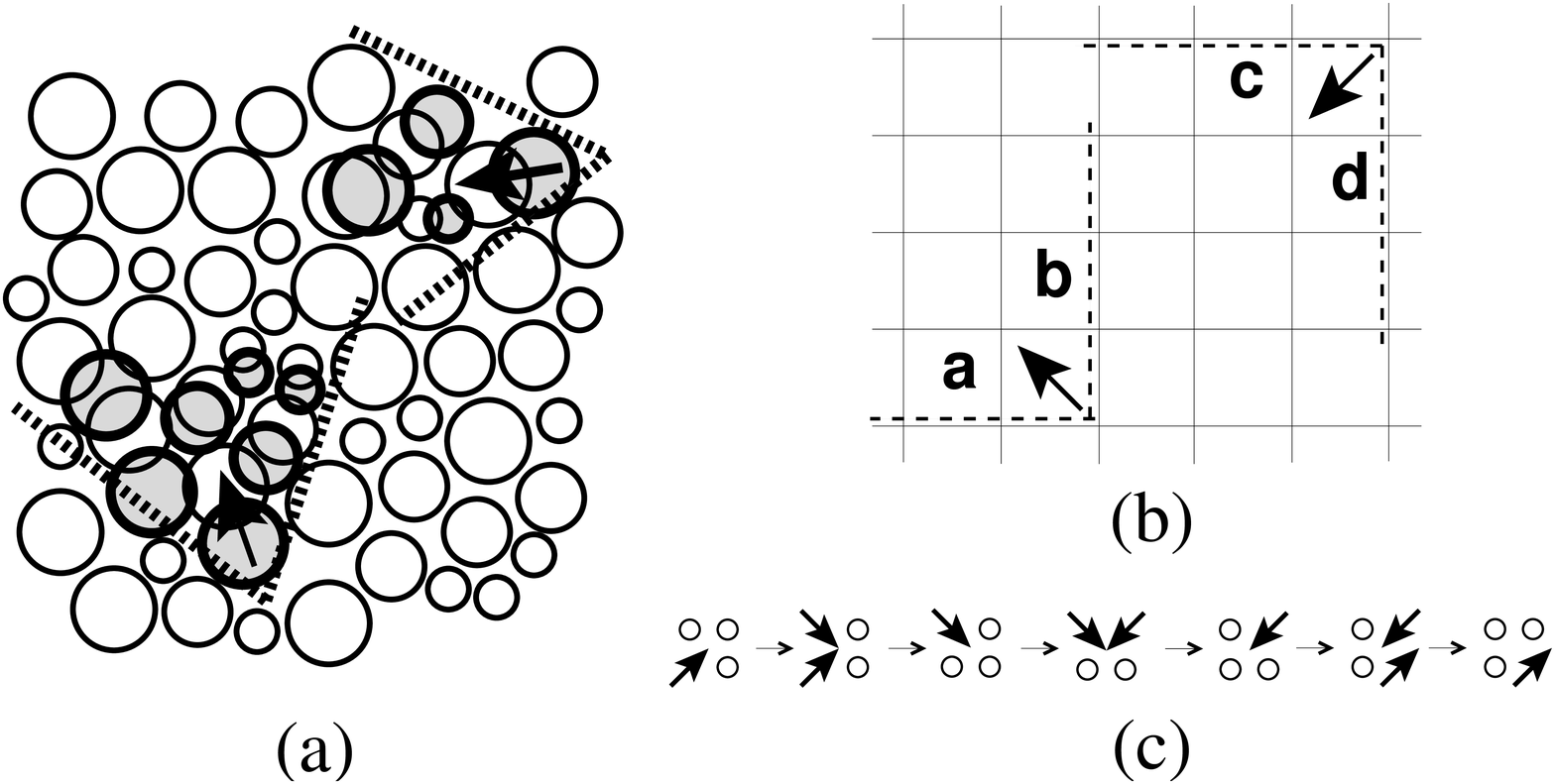, width=8.4cm}
\caption{(a) Facilitation of motion in a nearly arrested fluid.  The
arrows indicate direction of facilitation, anti parallel to direction
of flow.  Motion is facilitated in cones indicated by dotted lines.
The white circles depict positions of atoms at one point in time; each
grey circle, near a white circle of the same size, indicates the
position of that near atom at a later point in time.  (b) Coarse
grained lattice description. The excitation at the bottom facilitates
cells {\bf a} and {\bf b}, the excitation at the top, cells {\bf c}
and {\bf d}.  (c) Possible series of steps leading to a diffusive step
in $d=2$.}
\end{center}
\end{figure}

To capture these physical features, we use the coarse grained
description depicted in Fig.\ 1(b). Space is divided into square or
cubic regions with side lengths comparable to, but not smaller than,
the bulk correlation length of the fluid. As such, equilibrium
fluctuations in one cell are uncorrelated with those in another at the
same point in time.  Further, time is divided into discrete intervals
or time steps of length $\delta t$ long enough to discriminate whether
mobility (i.e., significant particle displacement) has or has not
occurred in a particular cell during that interval. An empty cell in
Fig.\ 1(b) indicates a region that exhibits no mobility during that
time step. A cell containing an arrow indicates a region that does
exhibit mobility during that time step, and the direction of the arrow
is anti-parallel to the direction of motion and therefore indicates
the direction of facilitation.

This coarse grained catalog or description of dynamical behavior can
be constructed from the trajectories of any system. In this sense,
there is no approximation involved in its use. Were the system an
ordinary liquid, a typical equilibrium configuration of the coarse
grained model would have essentially all regions occupied. For a
sufficiently supercooled liquid, however, the concentration of arrows
will be low. In that case, the thermodynamics of the material should
be consistent with that of non-interacting arrows on a lattice. As a
simplification, we shall consider only discrete orientations of the
arrows, restricting them to lie along any one of the principal
diagonals of its cell.  Placing arrows along cell diagonals rather
than edges allows for a cone of influence with positive angle.  In
particular, an arrow at a specific cell and time step can facilitate
the creation or destruction of an arrow in a nearest neighbor cell at
the next time step, where the vector drawn from the specific cell
towards that nearest neighbor has a positive inner product with the
arrow.  For example, in Fig.\ 1(b), the arrow at the lower left can
facilitate the birth of an arrow at the next time step in either cell
labeled {\bf a} or {\bf b}.  In this case, the cone of influence spans
a quadrant in the square lattice, or an octant in the cubic lattice.
Different cones of influence can be defined with different choices of
spatial grids.

\bigskip

\noindent
{\bf The model.} As the preceding discussion indicates, our model
describes the dynamics of the vector field ${\bf n}(\vec x)=n(\vec
x){\bf v}(\vec x)$, where $n(\vec x)=1$ or $0$, depending on whether
atoms in cell $\vec x$ are mobile or immobile, respectively, and ${\bf
v}(\vec x)$ is the unit vector pointing in the direction of the arrow
in cell $\vec x$. For a square ($d=2$) lattice, ${\bf v}(\vec x)$ will
be one of the four vectors $\left( \pm 1,\pm 1\right)/\sqrt{2}$. The
equilibrium concentration of mobile cells is denoted by
$c=\left\langle n(\vec x)\right\rangle$ (taking the lattice spacing to
be unit length). The pointed brackets indicate equilibrium ensemble
average. The system is isotropic, so that $\left\langle {\bf v}(\vec
x)\right\rangle =0$, and only trivial equal-time correlations exist so
that $\left\langle {\bf n}(\vec x){\bf n}(\vec y)\right\rangle=0$ for
$\vec x \neq \vec y$. Accordingly, the equilibrium distribution for
the vector field is $P\left( \left\{ {\bf n}(\vec x)\right\} \right)
=\prod_{\vec x} \rho\left[{\bf n}(\vec x)\right]$, where
\begin{equation}
\rho \left( {\bf n}\right) \equiv g^{-n} \left( 1-c\right)
^{1-n}\,c^{n} .
\label{distribution}
\end{equation}
Here, $g$ is the number of equally likely equilibrium orientations of
an arrow [i.e., a vector ${\bf v}(\vec x)$]. For the square (cubic)
lattice, $g=4$ $(g=8)$.

The passage from one time interval, $t$, to the next, $t+\delta t$, is
governed in our model by the following single site Monte Carlo
dynamics.  Consider the transition at site $\vec x$ of the vector
field ${\bf n}(\vec x)=n(\vec x){\bf v}(\vec x)$ from an unexcited
state, $n(\vec x)=0$, to an excited state $n(\vec x)=1$, ${\bf v}(\vec
x)={\bf w}$, and the corresponding backward transition.  The dynamic
facilitation described above is implemented by defining the
corresponding rates as:
\begin{equation}
{\bf n}(\vec x)=0 
\begin{array}{c}
\xrightarrow{~{\cal C}_{\vec x}[{\bf w}] ~ c/g~} \\
\xleftarrow[{\cal C}_{\vec x}{[{\bf w}]} ~ (1-c)]{} \\
\end{array}
{\bf n}(\vec x)={\bf w} ,
\label{rates}
\end{equation}
where the kinetic constraint is implemented by the indicated state
dependent rates, with ${\cal C}_{\vec x}{[{\bf w}]}$ given by
\begin{eqnarray}
{\cal C}_{\vec x}[{\bf w}] &=& f \left[ 1 - \prod_{\langle {\vec
y},{\vec x} \rangle} \left( 1-\delta_{\sqrt{d} (\vec x - \vec y) \cdot
{\bf n}(\vec y), 1} \right) \right] \nonumber \\ && + (1 - f
)\left[1-\prod_{i=1}^d \left( 1-\delta_{{\bf n} \left( \vec x -
\sqrt{d} {\bf w}_i \right), {\bf w}} \right)\right] .
\label{constraint}
\end{eqnarray}
Here, ${\bf w}_i$ indicates the $i$-th Cartesian component of ${\bf
w}$, nearest neighbors are denoted by $\langle {\vec y},{\vec x}
\rangle$.  Since ${\cal C}_{\vec x}{[{\bf w}]}$ does not depend on the
state of the cell making the transition, but only on the state of its
neighbours, it disappears from the ratio between forward and backward
rates, and the dynamics defined by (\ref{rates}) satisfies detailed
balance with respect to the distribution (\ref{distribution}).  The
parameter $f \in [0,1]$ in Eq.\ (\ref{constraint}) determines the
probability $p(f)$ that a newly created arrow will lie parallel to its
facilitating arrow, $p(f)=[1+f(g-1)]^{-1}$.  One of the extremes,
$f=1$, corresponds to the case where an excitation pointing in any
direction can be created or destroyed as long as its site is
facilitated by a neighbour.  In the other extreme, $f=0$, only
excitations pointing in the same direction as the facilitating
neighbour can be created or destroyed.

The above dynamics satisfies time reversal symmetry and detailed
balance.  It is designed to be physically realistic for $c$ small, the
realm of dynamical arrest in a glass former.  When $c$ is not small,
the dynamics of a realistic system will proceed through other
mechanisms in addition to facilitation.  In general, even for $c$
small, space and time coarse graining of an atomistic model will lead
to transition rules with non-Markovian elements. We assume these
elements are unimportant.

The parameters $c$ and $f$ have clear physical meaning.  Observations
of trajectories can determine the probability, $c$, that a microscopic
region exhibits mobility over a specified microscopic time period.
They can further determine $f$ from the directional persistence
probability over a similar time period.  For systems where
directionality in particle motion can persist over a microscopic time
$\delta t$ (e.g., a nearly jammed fluid of hard spheres), $f$ will be
very small.  For systems where directionality in
particle motion will dissipate over that time (e.g., a highly
networked fluid), $f$ will be close to $1$.  One may use atomistic
computer simulations of supercooled fluids to make the observations
that determine $c$ and $f$.  Alternatively, one may perform microscopy
on colloidal glass formers.  In these cases, observed time scales
relative to those of an ordinary liquid are modest in length.  For
glass forming supercooled liquids exhibiting long relative times
scales, experiments using single molecule spectroscopic probes could
be used.

The parameters $c$ and $f$ have energetic and entropic contributions.
Since excitations in different sites are statically uncorrelated,
their average concentration $c$ must have the usual Boltzmann
temperature dependence.  Therefore, to the extent that $c$ is small
and mainly energetic, $-\ln \left( c/g\right)$ is proportional to
reciprocal temperature, $1/T$.  Similarly, since $f$ is a local
probability independent of the state of the system, it is natural to
write $f=ac^{b}$ where $a$ pertains mainly to its entropic
contribution, and $b$ pertains mainly to its energetic contribution.

\bigskip

\noindent
{\bf Properties of the model}. It is useful to begin by considering
the two extreme limits of the model, the $f=1$ case and the $f=0$
case.

$\bullet\ f = 1$. Here, motion of an arrow can proceed diffusively as
illustrated with the seven step sequence configurations shown in Fig.\
1(c). This sequence, involving single activation on three occasions,
produces a single diffusive step in either $d=2$ or $3$. As such, the
relaxation time for a length scale $l$, $\tau_{f=1}(l)$, will obey
$\tau_{f=1}(l) \sim \delta t\,D^{-1}\,l^{z_1}$, where the dynamical
exponent, $z_1$, will be a constant, while the diffusion constant,
$D$, will be proportional to $c$ for $d=2$ and $3$. (The quantity $D$
is the diffusion constant for excitations, which is distinct from the
diffusion constant for particle motion.)  Since the distance between
arrows at equilibrium is $c^{-1/d}$, the system approaches equilibrium
with time scale $\tau_{f=1} \sim \delta t\,c^{-\Delta_1}$ where for
$d=2$ and $3$, $\Delta_1=1+z_1/d$.  From numerical simulations of the
model on a square ($d=2$) and cubic ($d=3$) lattices, we find that
$\Delta_1 \thickapprox 2$. Simulation results for the relaxation of
the model from a disordered state on a square lattice are shown in the
left panel of Fig.\ 2.  As expected from the diffusive nature of its
motion, the behavior of this model is like that of the FA (i.e.,
one-spin-facilitated Fredrickson-Andersen
\cite{Fredrickson-Andersen,Ritort-Sollich}) model. The Arrhenius
dependence of $\tau_{f=1}$ on $c$ is often called ``strong''
\cite{Review}.

$\bullet\ f = 0$. Here, the motion is hierarchical
\cite{Palmer-et-al,Garrahan-Chandler}, like that of the East model
\cite{East-model,Ritort-Sollich} or its generalization to higher
dimension \cite{North-East}, and it is ergodic for $d \geq 2$. An
arrow can relax only through pathways involving like pointing
arrows. For $d=2$ the system is composed of a mixture of arrows of
four different directions, with each direction uncoupled from the
others. The dynamical behavior of each of the four independent
components is like that of a North-or-East model \cite{North-East}
which in turn behaves like the East model. In this case, therefore,
$\tau_{f=0}(l) \sim \delta t\,l^{z_0}$ where now the dynamical
exponent is not constant but rather a function of $c$, namely $z_0
\sim \ln(g/c)/\ln 2$, where $c/g$ is the concentration of like
directed arrows. Hence, close to equilibrium, the overall time scale
for relaxation is $\tau_{f=0} \sim \delta t \, c^{-\Delta_0}$, where
$\Delta_0 = z_0/d$. This scaling is consistent with the simulation
results for $d=2$ graphed in the right panel of Fig.\ 2. The
super-Arrhenius dependence on $c$ is often called ``fragile''
\cite{Review}.

$\bullet\ 0<f<1$. Here, we adopt an interpolation formula by adding
rates in series: $\tau^{-1}\left(
l\right)\thickapprox(1-f)~\tau_{f=0}^{-1}\left( l\right) + f~
\tau_{f=1}^{-1}\left( l\right)$. As such, for small $f$,
\begin{equation}
\tau \left( l\right) \thickapprox \delta t\,l^{z_{0}}\left( 1+ D ~ f ~
l^{\left( z_{0} -z_{1} \right)} \right) ^{-1} .
\label{tau(l)}
\end{equation}
The corresponding equilibrium relaxation time is therefore
\begin{equation}
\tau \thickapprox \delta t \, c^{-\Delta _{o}} \left( 1+ f ~
c^{-(\Delta _{0}-\Delta _{1})} \right)^{-1} . \label{tau}
\end{equation}
These equations exhibit a crossover behavior. In particular, for a
given $c$ and $f$, there is a crossover length, $l_*$, such
that for $l>l_*$, relaxation is diffusive and thus Arrhenius,
and for $l<l_*$, relaxation is hierarchical and
super-Arrhenius. Thus, at large enough length scales, the relaxation
of all systems, no matter how fragile, will obey diffusive behavior.
The existence of the crossover length, $l_*$, could be
demonstrated experimentally from the single molecule spectroscopy of
rotating nanoparticle probes of various lengths. The existence of a
crossover length also implies that there is a crossover $c_*$
(or $T_*$ if one associates $c$ with $T$ as suggested earlier)
such that for $c<c_*$ (or $T<T_*$), the equilibrium
relaxation time of the model is Arrhenius, and for $c>c_*$ (or
$T>T_*)$, it is super-Arrhenius. Implicit in these remarks is
an appreciation for the presence of dynamic heterogeneity
\cite{Garrahan-Chandler,DHReviews}, in this case including flow
direction, but for limitations of space, we cannot discuss it further
here.

\begin{figure}[h]
\begin{center}
\epsfig{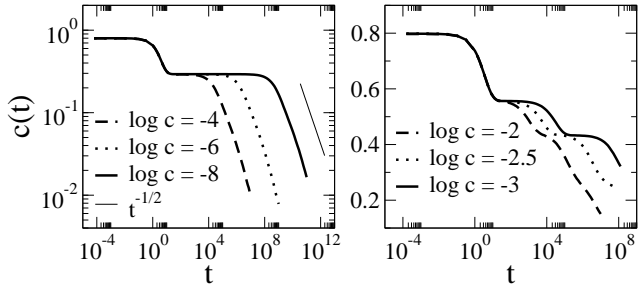}
\caption{Relaxation of the concentration $c(t)$ after a quench from an
initial random state to final equilibrium concentration $c$ in the
$d=2$ version of the model.  Left panel: $f=1$. Right panel: $f=0$.
Simulations were performed using continuous time Monte-Carlo
\cite{Newman-Barkema}, on system sizes $L=10^5$ and averaged over 50
samples.  Error bars are comparable to line thickness.  }
\end{center}
\end{figure}

Crossover behavior was already demonstrated for what can be considered
a mean field version of our model \cite{Buhot-Garrahan}. The presence
of a crossover has also been argued for on experimental grounds
\cite{Ito-Moynihan-Angell}.  Figure 3 illustrates the consequent
temperature dependence for the possible crossover behaviors predicted
from our model.  Graphs on the left are drawn as Arrhenius
plots. Graphs on the right show the same quantities but scaled in the
fashion proposed by Angell \cite{Review}. Notice the possibility that
relaxation times may fall above the Arrhenius line in the Angell
plot. This specific behavior has not been anticipated. It arises in
systems where $f$ is both small and mainly entropic (i.e., independent
of $c$).

\bigskip

\noindent
{\bf Comparison with experiment.} The fundamental control parameter is
$c$, and while this quantity could be measured directly, most existing
experimental data observes the effects of changing $c$ by changing
temperature at the fixed pressure of $1~$atm.  To compare with that
data, we therefore use
\begin{equation}
\ln \left( \frac{c}{1-c} \right) \thickapprox \ln \left( \frac{c_{{\rm
R}}}{1-c_{{\rm R}}} \right) - J ~ \left( \frac{1}{T} -
\frac{1}{T_{{\rm R}}} \right) ,
\label{cr}
\end{equation} 
where $c_{{\rm R}}$ is the concentration of mobile regions at a
reference temperature $T_{{\rm R}}$, and $J$ is the enthalpy
difference between a mobile and immobile region.  Equation (\ref{cr})
is again based on the idea that since excitations at different spatial
points (at equal times) are statically uncorrelated their average
concentration $c$ must have Boltzmann temperature dependence.  Since
mobile regions should be liquid like, and immobile regions should be
solid like, we expect that $J/T$ should be of the order of a few
entropy units times the typical number of molecules in a cell $(\sim
10)$. With $\ln c$ linear in $1/T $, $\ln \tau _{f=0}$ is a quadratic
polynomial in $T^{-1}$ \cite{Bassler}, while $\ln \tau _{f=1}$ is a
linear one.  A useful reference point, which allows for a direct
comparison with previous analysis (eg., Ref.\ \cite{Martinez-Angell}),
is $T_{{\rm R}}=T_{1/2}$, the temperature at which the logarithm of
the relaxation time is one half that at the glass transition. In this
case $\tau _{{\rm R}}\equiv \tau (c_{{\rm R}})$ is of the order of
microseconds $(\mu {\rm s})$. A common measure of fragility is
$F_{1/2}\equiv 2T_{{\rm g}}/T_{1/2}-1$, where $T_{{\rm g}}$ is the
glass transition temperature \cite {Martinez-Angell}. A second common
measure is the steepness index, $m\equiv d\log \tau (T_{{\rm
g}})/d(T_{{\rm g}}/T)$ \cite{mData}, i.e., the slope at $T_{{\rm g}}$
in the Angell plot. If the relaxation from $T_{1/2}$ to $T_{{\rm g}}$
follows the super-Arrhenius form of $\tau _{f=0}$, then $F_{1/2}$ and
the elementary timescale $\delta t$ determine $m$:
$m_{f=0}=32(1-F_{1/2})^{-1}(1+Y)(1+2Y)^{-1}$, where $Y=\log (\mu {\rm
s}/\delta t)\,[1+\sqrt{1+4/\log (\mu {\rm s}/\delta t)}]/8$. In the
case of Arrhenius behaviour, or of an energetic crossover that takes
place at $T>T_{{\rm g}}$, the steepness index is given by
$m_{f=1}=16(1-F_{1/2})^{-1}$, independent of $\delta t$.

\begin{figure}[h]
\begin{center}
\epsfig{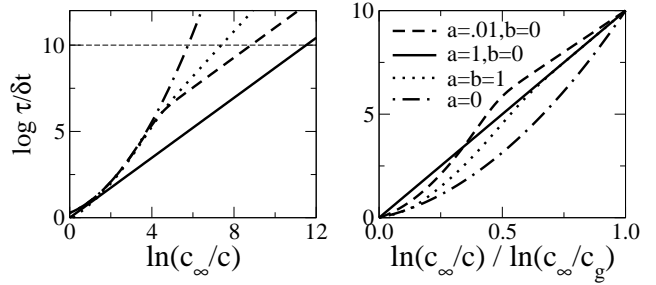}
\caption{Left panel: Arrhenius-like plot of relaxation time against
equilibrium concentration, for various values of $f=a \, c^b$. The
cases $a=1,b=0$ and $a=0$ correspond to purely Arrhenius and
super-Arrhenius behaviour, respectively. The case $a=b=1$ leads to an
energetic crossover, while $a=.01,b=0$, to a mainly entropic one.
Right panel: same as before now presented as an Angell-like plot, with
the glass transition chosen at $\tau \sim 10^{10} \delta
t$. $c_{\infty}=8/9$ is the random concentration value for the cubic
lattice.  }
\end{center}
\end{figure}

Figure 4 shows that the ranges of fragility measures
\cite{mData,Martinez-Angell} are well understood with our model.  The
dashed line corresponds to the value of $m_{f=0}$ in the case where
$\delta t\sim \mu s$, $m_{f=0}=32(1-F_{1/2})^{-1}$.  The full line is
$m_{f=1}$.  The region between this two lines, region I in Fig.\ 4,
corresponds to super-Arrhenius behaviour $\tau _{f=0}$ with $\delta
t\leq \mu s$.  Most glass forming liquids fall in this region. Liquids
on the full line are either those with purely Arrhenius behaviour or
those that undergo an energetic crossover. Region II below the full
line, corresponds to liquids which experience the entropic crossover
noted above.  Salol (SL) and $\alpha$-phenyl-$o$-cresol (APOC) fall
in this region.  Region III above the dashed line corresponds to
values of $m$ which are too large to account with the present approach
(except with $\delta t\gg \mu {\rm s}$, which seems unphysical).
Since selenium forms polymer chains, a feature not accounted for in
our model, it is perhaps not surprising that it falls in this region
\cite{selenium}.

In Fig.\ 5 we show Angell plots in the regime near $T_{{\rm g}}$ for
five representative liquids from Fig.\ 4 (data from \cite{data}). It
shows fits to the dielectric relaxation timescale of 3-bromopentane
(3BP) and viscosity of ortho-terphenyl (OTP), using $\tau _{f=0}$, and
an Arrhenius fit to the viscosity data of GeO$_{2}$, using
$\tau_{f=1}$.  It also illustrates the crossover in the the viscosity
data of APOC and SL.  The fits to the kinetic data determine the
parameters of the model. From the coefficients of the terms quadratic
in $T_{{\rm g}}/T$ in $\log \tau _{f=0}$, and from the linear one in
$\log \tau_{f=1}$, we obtain the values of $J/T_{{\rm
g}}=(16.7,26.7,29.3,31.8,22.6)$ for (3BP, OTP, APOC, SL,
GeO$_{2}$). These values of $J/T_{{\rm g}}$ obtained from fitting
kinetic data agree with the expectations of our model.

\begin{figure}[h]
\begin{center}
\epsfig{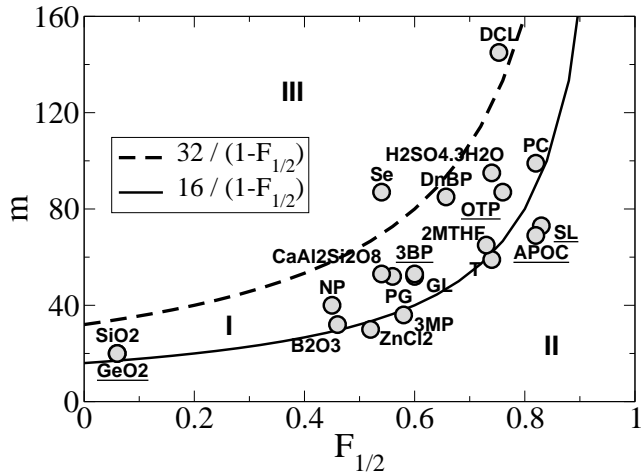}
\caption{Steepness index $m$ against kinetic fragility $F_{1/2}$ for
various supercooled liquids.}
\end{center}
\end{figure}

Further, from the linear and constant terms in $\log \tau _{f=0}$, we
determine $\log (\delta t/\mu {\rm s})=(-3.3,-4.4,-4.7,-4.8)$ and
$\log (c_{{\rm g}}/g)=(-3.2,-3.4,-3.6,-3.9)$, for (3BP, OTP, APOC,
SL), respectively, so that in these cases $\delta t$ is of the order
of fractions of ns. These are physically reasonable times to discern
whether or not a microscopic region exhibits mobility. We also obtain
a relation between the concentration at $T_{1/2}$, to that at the
glass transition, $\log \{c_{{\rm g}}(1-c_{1/2})/[c_{1/2}(1-c_{{\rm
g}})]\}=(-1.5,-1.4,-1.5,-1.8,-4.6)$, for (3BP, OTP, APOC,
SL,GeO$_{2}$). This list of ratios could in principle be checked
experimentally. Fits to kinetic data do not determine the value of the
parameter $g$. If we assume $g\thickapprox 8$ as in the lattice model,
then we obtain for the previous liquids: $\log c_{{\rm
g}}=(-2.3,-2.5,-2.7,-3.0,-3.6)$, where in the case of GeO$_{2}$ we
have taken $c_{1/2}\sim O(1)$. Notice that the mobility concentration
at $T_{{\rm g}}$ of the two purely super-Arrhenius liquids, 3BP and
OTP, is at least an order of magnitude larger than that of the purely
Arrhenius one, GeO$_{2}$. Moreover, the two liquids which display a
crossover have intermediate values of $c_{{\rm g}}$.

The order of magnitude difference between fragile and strong liquids
(OTP and 3BP versus GeO$_2$ in our example) in concentration of
mobility excitations at $T_{\rm g}$ is a crucial theoretical
prediction of our approach. In the context of the model this
difference has a clear origin.  Relaxation time grows with decreasing
$c$ much faster in the fragile case ($f=0$) than it in the strong case
($f=1$).  Thus, an arbitrarily specified long relaxation time (e.g.,
some minutes) coinciding to a glass transition is reached at a value
of $c$ in the fragile case that is much larger than that in the strong
case.

\begin{figure}[h]
\begin{center}
\epsfig{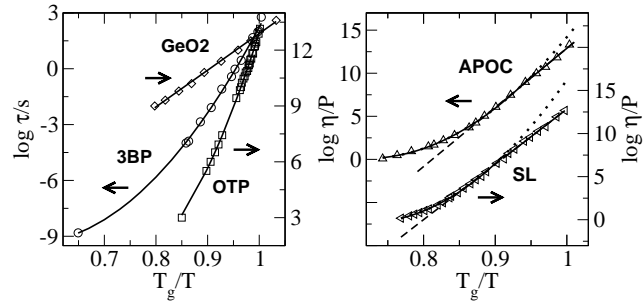}
\caption{Angell plots of dielectric relaxation and viscosity. In the
left panel, solid lines for 3BP and OTP refer to $\tau_{f=0}$, and
solid line for GeO$_2$ refers to $\tau_{f=1}$. In right panel full
line corresponds to Eq.\ (\ref{tau}); dotted and dashed lines indicate
$\tau_{f=0}$ and $\tau_{f=1}$ behaviour before and after the
crossover. Symbols indicate experimental data.  }
\end{center}
\end{figure}

The phenomenological consequences of this difference in concentrations
are important.  For example, $1/c^{1/d}(T)$ sets the dynamic
correlation length, which in turn gives an upper bound for the size of
dynamical heterogeneities \cite{Garrahan-Chandler}.  Assuming that
each cell contains about 5 to 10 molecules, so that $\delta x \approx
1$--$1.5$ nm, we estimate that the size of dynamic heterogeneities for
3BP, OTP and GeO$_2$ are bounded from above by $5$--$8$ nm, $6$--$9$
nm and $15$--$24$ nm, respectively.  As we have stressed
\cite{Garrahan-Chandler}, dynamic heterogeneity is self-similar over a
range of lengths.  Therefore, at a specific thermodynamic state point,
no single length scale can be attributed to dynamic heterogeneity.
Nevertheless, experimentalists have reported results for a few fragile
liquids as if there is but a single specific length
\cite{sizeheter,DHReviews}.  The reported experimental lengths are
consistent with our predicted bounds, though a definite comparison
between theory and experiment must await more complete experimental
information on the full range of heterogeneity lengths.  Since
$1/c_{\rm g}$ is larger in the strong case than the fragile case, our
model predicts that the size of dynamical heterogeneous regions in
strong liquids should be appreciably larger than those in fragile
liquids.  This predicted trend is opposite of that made by other
theoretical approaches, like frustrated limited domains
\cite{KivelsonTarjus} or random first order transitions
\cite{XiaWolynes}.

From the concentrations $c_{\rm g}$ and the corresponding values of
$J/T_{{\rm g}}$ we are able to predict the jumps in specific heats at
$T_{{\rm g}}$.  According to our model, this heat capacity (per mol)
relative to that of the solid is
\begin{equation}
\Delta c_{p}(T_{{\rm
g}})\thickapprox (J/T_{{\rm g}})^{2}\,c_{{\rm g}}\,{\cal N}+O(c_{{\rm
g}}^{2}), 
\end{equation}
where ${\cal N}$ is the number of molecules that contribute to
enthalpy fluctuations per mobile cell. With the values obtained from
the fits to the kinetic data for OTP, 3BP and GeO$_{2}$, for example,
we obtain, $\Delta c_{p}(T_{g})/{\cal N}\sim 2.3,1.3,0.13$,
respectively.  These values are in the ratio $17:10:1$, which compares
well to the experimental values $\Delta c_{p}(T_{g})/k_{{\rm B}}\sim
13.6,9.2,0.84$ \cite{cp}, whose ratio is $16:11:1$.

Further comparisons between this theory and experiment illustrating
subtle counter examples to the usual correlation between
super-Arrhenius behaviour and $\Delta c_p$ \cite{Huang-McKenna} are
left to future papers. Along with comparisons to equilibrium
properties, illustrated herein, our model can be applied to
out-of-equilibrium behaviors without further assumptions.

\bigskip

\noindent 
{\bf Acknowledgments.}  We are grateful to C.A.  Angell for
discussions, and to H.C. Andersen, D. Reichman, F. Ritort, and
F.H. Stillinger for discussions and constructive comments on an
earlier version of this paper.  This work was supported at Oxford by
the Glasstone Fund and EPSRC Grant No.\ GR/R83712/01, and at Berkeley
in its early stages by the NSF, and at its later stages by the US
Department of Energy grant DE-FG03-87ER13793.

\end{document}